\documentclass{pasj01}

\newcommand{\kms}{km s$^{-1}\;$}
\newcommand{\kmss}{km s$^{-1}$}

\newcommand{\msun}{\mbox{M$_{\odot}$}}
\newcommand{\lsun}{\mbox{L$_{\odot}$}}
\newcommand{\ho}{H$_{2}$O$\;$}

\newcommand{\nh}{$N$$_{\rm H}\;$}
\newcommand{\nhh}{$N$$_{\rm H}$}

\begin{document} 
\Received{2017/01/10}
\Accepted{2018/03/28}

\title{Searches for H$_2$O Masers toward Narrow-Line Seyfert 1 galaxies}

\author{Yoshiaki \textsc{hagiwara}\altaffilmark{1,2}}%
\altaffiltext{1}{Natural Science Laboratory, Toyo University, 5-28-20, Hakusan, Bunkyo-ku, Tokyo 112-8606, Japan}
\altaffiltext{2}{National Astronomical Observatory of Japan, 2-21-1, Osawa, Mitaka, 181-8588 Tokyo, Japan}
\email{yhagiwara@toyo.jp}

\author{Akihiro \textsc{doi}\altaffilmark{3}}
\altaffiltext{3}{The Institute of Space and Astronautical Science, Japan Aerospace Exploration Agency, 3-1-1, Yoshinodai, Chuou-ku, Sagamihara, Kanagawa 252-5210, Japan}

\author{Kazuya \textsc{hachisuka}\altaffilmark{4}}
\altaffiltext{4}{Mizusawa VLBI Observatory, National Astronomical Observatory of Japan, Hoshigaoka 2-12, Mizusawa, Oshu, Iwate 023-0861, Japan}

\author{Shinji \textsc{horiuchi}\altaffilmark{5}}
\altaffiltext{5}{CSIRO Astronomy and Space Science, Canberra Deep Space Communications Complex, PO Box 1035, Tuggeranong, ACT 2901, Australia}


%
\maketitle
\begin{abstract}
We present searches for 22 GHz \ho masers toward 36 narrow-line Seyfert 1 galaxies (NLS1s), selected from known NLS1s with $v_{sys}$ ~$\lesssim$ 41000 \kmss.
 Out of the 36 NLS1s in our sample, 11 NLS1s have been first surveyed in our observations, while the observations of other NLS1s were previously reported in literature. In our survey, no new water maser source from NLS1s was detected at 3 $\sigma$ rms level of 8.4 mJy to 144 mJy, which depends on different observing conditions or inhomogeneous sensitivities of each observation using three different telescopes. It is likely that non-detection of new masers in our NLS1 sample is primarily due to insufficient sensitivities of our observations. 
Including the five known NLS1 masers, the total detection rate of the \ho maser in NLS1s is not remarkably different from that of type 2 Seyfert galaxies or LINERs. However, more extensive and systematic searches of NLS1 would be required for statistical discussion of the detection rate of the NLS1 maser, compared with that of type 2 Seyferts or LINERs.
\end{abstract}

\section{Introduction}
Luminous extragalactic \ho masers in the transition of $6_{16}-5_{23}$ (rest frequency: 22.23508 GHz) are known to exist in central regions of active galactic nuclei (AGN).  
Many attempts at finding extragalactic \ho masers have been made, with the result that detections of these masers are largely in type 2 Seyfert galaxies and LINERs  (e.g., \cite{cla86,bra96,hagi02,kon06a,wag13}). 
Some fraction of these \ho masers associated with AGN-activity ("nuclear masers") have proven themselves to be useful in tracing the angular distribution of the maser spots within  about one parsec from a central engine. The velocity range of highly Doppler-shifted (high-velocity) maser emission in the nuclear masers is offset by up to $\sim$ $\pm$ 1000 \kms from galaxy's systemic velocity, which indicates presence of a rotating disc around a super massive black hole in host galaxies  \citep{miyo95,linc96,her99,hagi01,ish01,linc03b,kon05,kon08,mam09,rei09,kuo11,gao17}.

Narrow-line Seyfert 1 galaxies (NLS1s) were first studied by \citet{ost85}. These galaxies show the following optical spectral properties: 
1) The full width half maximum (FWHM) of the H$\beta$ lines is less than 2000 \kmss, 2) the permitted lines are somewhat broader
than the forbidden lines, and 3) the relative weakness of the forbidden lines, such as [O$_{~\rm{III}}$]$\lambda$5007 / H$\beta$ $<$ 3 \citep{ost85,goo89}.
There is evidence that NLS1s have mass accretion rates much closer to the Eddington limit than normal broad-line Seyfert galaxies and the NLS1s accreting near Eddington limit could have smaller black hole masses ($\sim$ 10$^6~$\msun) \citep{bor92,bol96,mine00,pet00}. The smaller masses of NLS1s are still under debate; their significantly smaller masses than expected from the M$_{BH}$ -- $\sigma$ relation have been shown compared to broad-line AGNs and quiescent galaxies (e.g., \cite{mat01,gru04}), while no strong evidence of such a deviation has been presented (e.g., \cite{wan01,kom08,woo15}). It is essential to understand the structure of NLS1's central engine with some different approaches.

\citet{hagi03} first discovered a bright \ho maser toward the narrow-line Seyfert 1 galaxy, NGC\,4051.
Stimulated by this discovery,  single-dish observations searching for new extragalactic \ho masers toward narrow-line Seyfert 1 galaxies were conducted, and these observations yielded the detections of new \ho masers in three narrow-line Seyfert 1 galaxies of Mrk\,766, IRAS\,03450+0055, and IGR\, J16385-2057 \citep{tar11}.
The detection of \ho maser emission toward a Seyfert 1.5 galaxy, NGC\,4151
was reported by \citet{bra04}: The galaxy is known to host a well-studied broad-line region and shows intermediate optical properties between type 1 and type 2 Seyferts (e.g., \cite{mun95}). These results show that \ho masers in active galaxies (a.k.a. ``megamasers'') have been found in type 2 Seyfert, LINER, narrow-line Seyfert 1, and other types of Seyfert nuclei. Some fraction of the megamasers are considered to be associated with ejecta from AGNs, such as jets or winds (e.g., \cite{cla98,linc03b}).
The recent study of radio-quiet NLS1s (e.g. Mrk\,1239, Mrk\,766) using very long baseline interferometry (VLBI) revealed that some NLS1s exhibit parsec-scale radio jets within 300 pc of the central engine \citep{doi13,doi15a,doi15b}.
The results would imply that \ho masers in NLS1s are a potential tracer of the circumnuclear regions
of AGN by analogy with the masers in type 2 Seyferts with (sub)parsec-scale non-thermal jets.
 The \ho masers in NLS1s provide important information about the geometry and kinematics of a disc or disc-like structure, jets and winds around the central engine of AGN, like the cases of the \ho masers in type 2 Seyferts, LINERs, and radio galaxies (e.g., \cite{miyo95,hagi01,linc03b,kuo11,ott13}). However, the origin of \ho masers in NLS1s has not been well understood, because most of them are not bright enough to be imaged at milliarcsecond (mas) angular resolution using VLBI.  

This article presents studies of \ho maser in NLS1s, based on
observations using the 45 m telescope at Nobeyama Radio Observatory
(NRO), the 100 m telescope of the Max-Planck-Institut f\"{u}r
Radioastronomy (MPIfR), and {the NASA Deep Space Network 70m telescope at Tidbinbilla (DSS-43)}. The studies of many \ho maser sources in NLS1s will
uncover radio properties of this class of AGN, and provide clues for
solving problems in the unified theory (e.g., \cite{ant85,urr95}). 
The article also uses results from other searches for statistics on maser detection.
Throughout this article, cosmological parameters of H$_{0}$ = 73 \kmss  Mpc$^
{-1}$, $\Omega$$_{\Lambda}$ = 0.73, and $\Omega$$_{M}$ = 0.27 are adopted.

\section{Sample Selection}
In our survey, relatively nearby 36 NLS1s ($v_{sys}$ $\lesssim$ 41000 \kmss) from NLS1 galaxy samples in  literature \citep{mat01,veron01,deo06,wha06} were programmed to search for new H$_2$O maser sources. Two NLS1s hosting known H$_2$O maser emission, NGC\,4051 and NGC\,5506 were selected to search for new high-velocity components; NGC5506 has been identified as an obscured narrow-line Seyfert 1 galaxy by near-IR spectroscopy \citep{nag02}, but not the definition based on optical spectral properties as summarized in the previous section.
The 26 NLS1s were observed in 2005-2006 at Nobeyama or Effelsberg. In addition, 14 NLS1s with relatively larger systemic velocities or lower declinations, of which 4 NLS1s were observed also at Nobeyama or Effelsberg, were observed in 2012 at Tidbinbilla (Table~\ref{tab0}).  In all, 36 NLS1s were observed in our program.
Mrk\,766 that implies an evidence for an accretion disc around the central black hole from X-ray observations \citep{tur06} was included to our list without knowing the fact that the galaxy was observed in other surveys.
Coordinates and systemic velocities of galaxies observed in our program, 1 $\sigma$ sensitivities, and velocity resolutions at each telescope are listed in Table~\ref{tab1}.

\section{Observations}
Single-dish searches for 22 GHz \ho maser emission toward NLS1s were made during the periods of 25 May to 3 June, 2005, 3 to 6 February, 2006, 12 to 14 March, 2012, and 23 July to 23 October, 2012. 
These observations in the first and third periods were carried out using the Nobeyama 45 m telescope (NRO 45 m), in the second period were conducted using the Effelsberg 100 m telescope, and in the fourth period were made using the Deep Space Network 70 m telescope at Tidbinbilla.
 All the observations in this program were made in the position-switching mode.  
 These observations are summarized in Table~\ref{tab0}.
 Originally, all 26 sources were scheduled at both Nobeyama 45 m and Effelsberg 100 m telescopes, however some sources were observed once and some were twice due to technical problems of telescope back-end at Effelsberg or weather conditions. In addition, 10 new sources below +20 degree declinations were scheduled at Tidbinbilla.
\subsection{Nobeyama 45 m}
The half-power beam width (HPBW) of the NRO 45 m was $\sim$74$\arcsec$ at 22 GHz and the system noise temperature was about 90 K to 160 K. The pointing calibration of each galaxy was conducted every 1h to 2h by observing 43 GHz SiO maser stars near galaxies,  which resulted in the pointing accuracy of $\sim$ 5$\arcsec$ --10$\arcsec$, typically.
A conversion factor of the antenna temperature to the flux density is estimated to be 2.63 Jy K$^{-1}$, adopting the aperture efficiency value of 66\% (T.Umemoto 2011 private communication), and flux density accuracy of $\sim$10\% is estimated.
In the Nobeyama 45m observations in 2005,  Acousto-optical spectrometer (AOS) was configured to record 2048 frequency points over a 40 MHz bandwidth for both left and right circularly polarized signals.
 We used an array of eight AOSs, resulting in a total velocity coverage of $\sim$ 2100 \kms for each polarization, nearly centred on the systemic velocity and having a frequency resolution of 39 kHz ($\sim$ 0.5  \kmss).  In the 2012 observations the new FX-type SAM45 spectrometer was used for the telescope back-end, in which an array of eight IFs subdivided into 250 MHz bandwidth was employed, each of them has 4096 spectral points, providing 61 kHz frequency resolution. The total velocity coverage and velocity resolution are 26000  \kms and 0.83 \kmss. In our analysis of the SAM45 data, two frequency channels were smoothed, which resulted in the velocity resolution of $\sim$ 1.6 \kmss. (Note that in Table~\ref{tab1} the sources listed with a 0.5  \kms velocity resolution were observed using the AOS spectrometer, and those with the 1.6 \kms resolution were observed using the SAM45.)
\subsection{Effelsberg 100 m}
The system temperatures and HPBW beam size were about 70 K to 120 K and $\sim$40$\arcsec$ in 22 GHz observations of the Effelsberg 100 m.  At Effelsberg eight-channel autocorrelators (AK90)
 were employed, each of them having a 40 MHz IF bandwidth and 512 spectral points, which yielded in 78 kHz frequency resolution. In our observations, four IF bands for each circular polarization were being used, which resulted in the total velocity coverage of $\sim$ 2100 \kmss, nearly centred on the systemic velocity and velocity resolution of $\sim$ 1.1 \kmss.
The pointing calibration was made by observing 
nearby strong continuum sources every 1h to 1.5 h, yielding 
a pointing accuracy of better than  $\sim$ 5 arcsec.    The telescope sensitivity of 3.1 Jy K$^{-1}$ is estimated, based on the standard gain curve formula of the Effelsberg 100 m  \citep{gal01}.  Accuracy of flux density is estimated to be $\sim$10\%.

\subsection{Tidbinbilla 70 m}

The system temperatures and HPBW beam size were about 25 K to 160 K and $\sim$48$\arcsec$.
Pointing errors were measured and corrected by using nearby bright quasars before observations, resulting a pointing accuracy of $\sim$ 7 $\arcsec$.  The Australian Telescope National Facility (ATNF) Correlator was configured to record 2048 spectral channels per polarization over a 64 MHz bandwidth for both left and right circularly polarized signals (e.g., \cite{sur09,bre13}), yielding a total velocity coverage of $\sim$ 860 \kms and spectral resolution of 31.25 kHz or 0.42 \kmss. The telescope sensitivity of 1.5 Jy K$^{-1}$ is adopted \citep{linc03a}, and accuracy of flux density is estimated to be $\sim$10\%.

Data reduction was conducted using the software packages NEWSTAR for NRO 45 m data, CLASS for Effelsberg 100 m data, and ATNF Spectral Analysis Package (ASAP) for Tidbinbilla 70 m data. Some data flagging was required due to spurious-like peaks or noises appearing in the band edges.

In this article, the velocities are calculated with respect to the Local Standard of Rest (LSR), and
 at Effelsberg and Tidbinbilla,  the optical convention is adopted, while the radio velocity definition is used at Nobeyama.

\section{Results} 
\label{sec:results}
In this program, no new H$_2$O maser source was detected at 3$\sigma$ detection level of 
$\sim$ 8 mJy -- 90 mJy per a spectral channel, by excluding the NRO 45 m observation of IRAS 17020+4544 that shows a very high rms value.
This implies that no strong H$_2$O maser was in the observed NLS1s during the observing periods.
 In our observations, the maser from Mrk\,766 was not detected, which is likely due to intensity variability of the maser. The maser in Mrk\,766 should have been marginally detected in 2$\sigma$ at Effelsberg, if its peak flux density was as bright as that observed at the NRAO Green Bank Telescope (GBT) in 2008 ($\sim$ 15 mJy) \citep{tar11}.
 {The H$_2$O  maser spectra of NGC\,4051 in Figure~\ref{f1} show the flux variability between two epochs of} the two known features at V = 679 and 738 -- 741 \kmss, which were reported in earlier observations \citep{hagi03,hagi07}. No new H$_2$O maser features were found in the observed bands toward the galaxy. 
  The maser emission in the NLS1s was searched in the velocity range of $\sim$ $\pm$ 400 -- 800 km s$^{-1}$ with respect to the systemic velocities of galaxies.  Measured rms noises were $\sim$ 9 mJy to 48 mJy per smoothed two or three channels (SAM45 or AOS) at Nobeyama, $\sim$ 3 mJy to $\sim$ 9 mJy at Effelsberg, and $\sim$ 4 mJy to $\sim$ 30 mJy at Tidbinbilla (Table~\ref{tab1}). We targeted 36 NLS1s \citep{ulv95,veron01,kom06,mul08}, out of which 26 NLS1s overlap those observed in the survey using the GBT \citep{tar11}. 
  In \cite{tar11}, new detections of the masers in two NLS1s, IGR J16385-2057 and IRAS 03450+0055 were reported. These two known NLS1 maser galaxies are not included in our sample.
\section{Discussion} 
\label{sec:discussion}
\subsection{No detection of new \ho maser in NLS1}
Of the 36 sources observed in our observations, Mrk\,766 is included. However, the maser in the galaxy was not detected both in 2005 and 2006 in our observations.  After our observations, the maser was detected in the NLS1 survey in 2008 with the GBT \citep{tar11}. It should be noted that the survey was more sensitive (1 $\sigma$ $\lesssim$ 3 mJy per a spectral channel of 0.33 \kmss) than ours (1 $\sigma$ $\sim$3 -- 48 mJy per a channel, or 7.3/12 mJy for Mrk 766).
Including the five known NLS1 masers, 71 NLS1s were surveyed with the GBT by Tarchi et al. and in our survey and 11 NLS1s were surveyed only in our survey, as a result of which the total detection rate of the \ho maser in NLS1s is estimated to be $\sim$6\% (5/82). 
It is shown that the nominal detection rates of the past \ho megamaser surveys of type 2 Seyferts and LINERs are $\sim$ 3\% \citep{bra97,bos16}, and in a survey of 40 AGNs with the GBT \citep{linc09},  the detection rate was 3/40 = 7.5\% in the $v_{sys}$ $<$ 20000 \kms sample. The GBT "snapshot" survey of nearby galaxies ($v_{sys}$ $<$ 5000 \kms) by \citet{bra08} has the detection rate of 1.3\%, however the survey detected
eight new masers.
Thus, the detection rate of the NLS1 maser is not remarkably different from that in previous extragalactic \ho maser surveys whose detection rates are, at most, several percent.
Moreover, the recent systematic study that compiled the results of past megamaser surveys revealed that the detection rate will be improved from $\sim$3\% to $\sim$16\% with the bias of higher extinction and higher optical luminosity, that is,  the luminosity of [O$_{\rm{III}}$] $\lambda$5007 \citep{zhu11}. 
Finally, we speculate that there is no evidence that incidence of \ho maser in NLS1s is either more or less probable than in other AGN masers in type 2 Seyferts or LINERs. 

One of the most critical problem in our programme lies in insufficient sensitivities of our survey, which makes it difficult to have statistical discussion, compared to past surveys. This is, for example, consistent with our non-detection of the Mrk\,766 maser.
\subsection{Column density and X-ray properties of \ho megamasers}
In earlier studies, correlation between the incidence of \ho maser emission and high column density ($N_{\rm H}$) in type 2 Seyferts and LINERs was demonstrated \citep{kon06a, zhan06, linc08, cas13}: Of 42 AGN megamasers whose column densities are available from published hard X-ray data, 95$\%$ have $N_{\rm H}$ $\gtrsim$ 10$^{23}$  cm$^{-2}$, or, alternatively, 60$\%$ are Compton thick, with $N_{\rm H}$ $\gtrsim$ 10$^{24}$  cm$^{-2}$ \citep{linc08}. 
Moreover, of 21 disc masers in which masers originate in subparsec- or parsec-sacle discs, 76$\%$ are Compton thick and the others are 10$^{23}$ cm$^{-2}$ $\lesssim$ $N_{\rm H}$ $\lesssim$10$^{24}$ cm$^{-2}$\citep{linc08}. 
According to the recent study with the X-ray observatory NuSTAR (Nuclear Spectroscopic Telescope Array) in the high-energy X-ray range (3 -- 79 keV), of 14 disc masers in type 2 Seyferts, 79$\%$ of masers are Compton-thick, and 21$\%$ are Compton-thin \citep{mas16}, which is largely consistent with 
the earlier study  by \citet{linc08}.  In contrast, the column densities of NLS1s are no higher than 10$^{24}$ cm$^{-2}$ (e.g., \citet{pan11}).
The summary of column density ($N_{\rm H}$) for the known NLS1 \ho maser is shown in Table~\ref{tab2}, in which there is no NLS1 showing column densities in excess of $N_{\rm H}$ $\sim$10$^{24}$ cm$^{-2}$, showing that  nuclear obscuration in NLS1s, including those hosting the maser, is smaller than those in type 2 Seyferts or LINERs.

Column densities toward active nuclei are obtained from \citet{mad06} (NGC\,4051), \citet{ris11} (Mrk\,766), and \citet{mol13} (NGC\,5506 and IGR\,J16385-2057), except for IRAS\,03450+0055 whose column density measured by hard X-ray toward a nucleus is not available in the literature (Table~\ref{tab2}). 
It is interesting to note that one of the best-studied disc masers, NGC\,4258 ($N_{\rm H}$ = 0.6 -- 1.3 $\times$ 10$^{23}$ cm$^{-2}$) and a known disc maser NGC\,4388 ($N_{\rm H}$ = 0.02 -- 4.8 $\times$ 10$^{23}$ cm$^{-2}$) are Compton thin,  \citep{mad06}, hence the occurrence of the maser cannot be explained simply by high column density. 
\subsection{Origin of \ho masers in NLS1s}
The small number of the detection of NLS1 masers demonstrates that masers are seen less enhanced because the masing discs are viewed less edge-on in line of sight, which is consistent with a picture of type 1 Seyfert nuclei in {AGN unified model \citep{ant85,urr95}}. 
The masers in NLS1s are associated with AGN-activity like other AGN masers. However, their averaged apparent maser luminosity (10 $\lsun$ $\lesssim$ $L(\rm{H_2O})$ $\lesssim$ 100 $\lsun$) is one order of magnitude lower than that of high-luminosity masers with $L{\rm( H_2O)}$ $\gtrsim$ 100  $\lsun$ \citep{hagi07}.  According to \citet{zhan06}, comparison between kilomasers ($L(\rm{H_2O})$ $\lesssim$ 10 $\lsun$) and high-luminosity masers shows that high-luminosity masers have higher \nhh.
Thus, it is less likely that a number of NLS1 masers with high luminosity will be detected in future single-dish surveys.  However, we expect to find as many NLS1 masers with luminosity as low (10 $\lsun$ $\lesssim$ $L(\rm{H_2O})$ $\lesssim$ 100 $\lsun$) as in other AGN masers.
Fig.~\ref{f2} shows the spectra of the three NLS1 masers, obtained by \citet{tar11}. The velocity ranges
 of the observed maser emission in these galaxies and NGC \,4051 are smaller than $\sim$ 100 \kmss, which could be
 explained by lower disc inclination angle of NLS1s: the apparent line of sight velocities (v$_{\rm{los}}$) are expressed as v$_{\rm{los}}$ = v$_{\rm{rad}}$ sin$i$, where $i$ is a disc inclination angle, and v$_{\rm{los}}$ with $i$ = 30$\arcdeg$--40$\arcdeg$ is calculated to be about 30--50\% smaller than those in more edge-on disc ($i$ $>$ 70$\arcdeg$). Alternatively, there is the possibility that these masers are associated with jets or winds in nuclear regions. We need a larger sample of the masers in NLS1s to examine these possibilities.
\subsection{NGC\,4051}
Fig.~\ref{f1} demonstrates that the maser flux density in NGC\,4051 is by a factor of $\sim$3 weaker than that in earlier observations due to variability. The maser in the galaxy was first detected in 2002 by \citet{hagi03}. They argued about the maser in the galaxy being a disc maser associated with an active nucleus. However, to date there has been no direct evidence that the maser originates in a parsec or sub-parsec scale disc around the nucleus in the galaxy. \citet{mad06} explains that broadly distributed narrow line emission in the galaxy is consistent with a wind maser associated with nuclear galactic winds related to formation of narrow-line Seyfert 1 spectra \citep{linc03b}. The galaxy exhibits significant variability in column density caused by the ionized absorbing medium \citep{mch95}, which should be physically separated from the molecular medium giving a rise to the maser, so the variability of the maser is not due to variability of the column density. Similarly, Mrk\,766 shows \nh variability that originates from ionized gas \citep{ris11} and not from the masing medium. The significant variability of the maser may be explained by the variability of background continuum in the nuclear region as in NGC\,6240 \citep{hagi15}, whereas 22 GHz nuclear continuum from NGC\,4051 has not been detected.
\section{Summary}
We searched for 22 GHz \ho masers toward 36 narrow-line Seyfert 1 galaxies (NLS1s) using the Nobeyama 45 m, Effelsberg 100 m, and Tidbinbilla 70m telescopes. 
We did not detect any new maser sources toward these NLS1s. We discussed possible causes of the non-detection, one of which is small number statistics by considering the overall detection rate of 3\% in previous extragalactic maser surveys and the other is insufficient sensitivities of our survey. There is no evidence for the occurrence of masers in NLS1s being higher or lower than in type 2 Seyferts or LINERs, although the higher detection rate of NLS1s masers is claimed in \citet{tar11}.

More detections of new maser sources in NLS1s would be necessary to establish the  overall nature of NLS1s masers. However, the number of detections of the maser sources in NLS1s could be small due to their low maser luminosity. 
We note that no high-velocity maser features in NGC\,4051 have been detected since the first detection in 2002. This might imply that the maser in the galaxy is not a disc maser but a wind maser associated with nuclear winds.

\begin{ack}
This research was supported by Japan Society for the Promotion of Science (JSPS) Grant-in-Aid for Scientific Research (B) (Grant Number: JP15H03644). Nobeyama Radio Observatory is a branch of the National Astronomical Observatory of Japan, National Institutes of Natural Sciences. Based on observations with the 100 m telescope of the MPIfR (Max-Planck-Institut f\"{u}r Radioastronomie) 
at Effelsberg. The Tidbinbilla 70m DSS-43 telescope is part of the Canberra Deep Space Communication Complex (CDSCC), which is operated by CSIRO Astronomy and Space Science on behalf of NASA. 
This research has made use of the NASA/IPAC Extragalactic Database (NED) which is operated by the Jet Propulsion Laboratory, California Institute of Technology, under contract with the National Aeronautics and Space Administration. YH thanks Koichiro Sugiyama and Teppei Tamaki for their help in observations at Nobeyama, and Alan Roy for useful comments to improve the initial draft of this article. YH also thanks Toshihiro Kawaguchi for valuable comments and suggestions.
\end{ack}
\onecolumn
{
\begin{center}
\begin{table}
\tbl{Summary of observations}{%
\begin{tabular}{lcccc} 
\hline
Telescope~~~~~~~~&Date& Number of sources$^{*}$ &Sensitivity (1$\sigma$) (mJy)& Note \\
\hline
Nobeyama  &  2005 May 25 -- June 3  & 20 &8.7--48&AOS spectrometer used    \\
  &  2012 March 12 --14  & 3 &15--24 &SAM45 spectrometer used     \\
Effelsberg  &  2006 February 3   --  6 & 19&2.8--8.5 &      \\
Tidbinbilla &  2012 July 23 -- October 23 &14& 4.3--27& Sources below +20 degree\\
&&&& declinations included  \\
\hline
\end{tabular}}\label{tab0}

Note: $^{*}$ The total number of targeted NLS1 galaxies is 36 as listed in Table~\ref{tab1}, 
while some of them were observed multiple times.
\end{table}
\end{center}
}

\begin{longtable}{lccrcccccc}
\caption{Summary of observations of narrow-line Seyfert 1 galaxies} \label{tab1}
\hline
Source$^{1}$~~~~~~~&${\alpha_{2000}}^{2}$&${\delta_{2000}}^{2}$&$V^{3}$~~~&$\sigma_{\rm {45}}$&$\sigma_{\rm{100}}$ &$\sigma_{\rm {70}}$& $\Delta\nu_{\rm{45}}$&$\Delta\nu_{\rm{100}}$ &$\Delta\nu_{\rm{70}}$\\
&&&(km s$^{-1}$)&(mJy)&(mJy)&(mJy)&(km s$^{-1}$)&(km s$^{-1}$)&(km s$^{-1}$) \\
\hline
\hline
\endfirsthead
\hline
Source~~~~~~~~~~~&$\alpha_{2000}$&$\delta_{2000}$&$V$~~~&$\sigma_{\rm {45}}$ &$\sigma_{\rm{100}}$ &$\sigma_{\rm {70}}$& $\Delta\nu_{\rm{45}}$&$\Delta\nu_{\rm{100}}$ &$\Delta\nu_{\rm{70}}$\\
\hline
\endhead
MRK\,335&00h06m19.5s&+20d12m10s &7730 & 21&4.0&&1.5&1.1 &\\
I\,Zw\,1&00h53m34.9s&+12d41m36s&18330&9.8& &6&1.5 & &1.6\\
MRK\,359& 01h27m32.5s& +19d10m44s&5212&17& & &1.5 & &\\
MRK\,1044&02h30m05.5s&-08d59m53s&4932 &15&&9&1.5& &1.6\\
MRK\,618&04h36m22.2s&-10d22m34s&10658& &&4.3&&&1.6\\
PKS\,0558--504   &05h59m47.4s   &  -50d26m52s  &  41132  & & &6.5&&&1.6\\
J\,07084153--4933066&07h08m41.5s   &  -49d33m07s  &    12162 & &&5.5&&&1.6\\
MRK\,382   &   07h55m25.3s   &  +39d11m10s  &      10139   &16&3.6 &&1.5&1.1&\\
                   &                            &                        &                    &  15&&&1.6&&\\
MRK\,0110   &   09h25m12.8s   & +52d17m10s   &      10580  & 17&5.3 &&1.5&1.1&\\
MRK\,705   &    09h26m03.3s   & +12d44m04s   &       8739   &8.7&3.3 &12&1.5&1.1&1.6\\
MRK\,0124   &          09h48m42.6s   & +50d29m31s   &       16878  & 15&2.9 &&1.5&1.1&\\
MRK\,1239   & 09h52m19.0s   & -01d36m43s   &       5974  &14&8.5 &5.5&1.5&1.1&1.6\\
KUG\,1031+398&      10h34m38.6s   & +39d38m28s   &      12724  &16& 3.8  &&1.5&1.1&\\
MRK\,734      &       11h21m47.0s   & +11d44m18s  &       15050   &&3.1  &&&1.1&\\
MRK\,42      &        11h53m41.8s   &  +46d12m43s  &       7385   &17& &&1.5&&\\
\bf{NGC\,4051}&  12h03m09.6s   &  +44d31m53s  &        700     &16&5.9&&1.5&1.1&\\
                         &                           &                         &               &24&&&1.6&&\\
\bf{MRK\,766} &  12h18m26.5s   &  +29d48m46s  &       3876   &12& 7.3 &&1.5&1.1&\\
                       &                          &                          &                   &15&  &    &1.6&&\\
WAS\,61   &         12h42m10.6s   &  +33d17m03s  &        13045     & &3.2 &&&1.1&\\
J\,12431152--0053442& 12h43m11.5s   & -00d53m45s  &24546&&& 31& &&1.6\\
NGC\,4748   &      12h52m12.4s   &  -13d24m53s  &     4386   & &  &5.0 && &1.6\\
MRK\,783 & 13h02m58.9s   &  +16d24m27s  &       20146   &21& 2.9&&1.5&1.1& \\
IRAS\,13224--3809&      13h25m19.4s   &  -38d24m53s  &     19726      && &22 && &1.6\\
PG\,1404+226    &     14h06m21.8s   &  +22d23m46s  &      29380     &17 &&&1.5&&\\
\bf{NGC\,5506} &       14h13m14.8s   &  -03d12m27s  &       1853     & & 5.6 &&  &1.1&\\
MRK\,478     &        14h42m07.4s   &  +35d26m23s  &      23700    & &3.4&&&1.1&\\
PG\,1448+273   &      14h51m08.8s   &  +27d09m27s  &      19487   &18& 6.5 &&1.5&1.1&\\
IRAS\,15091-2107 &      15h11m59.8s   &  -21d19m02s  &     13373     &&  & 12& & &1.6\\
15480--051      &  15h48m56.8s   &  -04d59m34s  &     29917   &&2.9&&&1.1 &\\
MRK\,493    &         15h59m09.6s   &  +35d01m48s  &       9392   &12& &&1.5&&\\
IRAS\,17020+4544 &    17h03m30.4s  &   +45d40m47s  &      18107  &48 & 3.0 &&1.5&1.1&\\
MRK\,507     &       17h48m38.4s  &   +68d42m16s   &     16758   &14& &&1.5&&\\
1927+654   &         19h27m19.5s  &  +65d33m54s   &      5096    & & 2.8&&&1.1& \\
1H\,1934--063\,A  &      19h37m33.0s   &  -06d13m05s  &     3074  & & & 27& & &1.6\\
2159+0113&      21h59m24.0s   &  +01d13m05s  &     30041  & &  &5.0& & &1.6\\
AKN\,564   &          22h42m39.3s  &   +29d43m31s  &       7400  &9.8&4.3 &&1.5&1.1& \\
2327--1023 &      23h26m56.1s   &  -10d21m43s  &     19557      & &  &4.6 & & &1.6\\ \hline 

\multicolumn{10}{l}{\small{$^{1}$ NLS1s hosting 22 GHz \ho maser are indicated in bold type.}} \\ 
\multicolumn{10}{l}{\small{$^{2}$ Right ascension ($\alpha_{2000}$) and declination ($\delta_{2000}$). }} \\ 
\multicolumn{10}{l}{\small{~~~Source coordinates are taken from NED.}} \\ 
\multicolumn{10}{l}{\small{$^{3}$ Systemic velocity (heliocentric definition from NED)}} \\ 
\multicolumn{10}{l}{\small{Note: $\sigma_{\rm {45}}$, $\sigma_{\rm {100}}$, and $\sigma_{\rm {70}}$ are 1 $\sigma$ noise per velocity resolution.}} \\ 
\multicolumn{10}{l}{\small{$\Delta\nu_{\rm{45}}, \Delta\nu_{\rm{100}},$ and $\Delta\nu_{\rm{70}}$ are velocity resolutions at Nobeyema,}} \\ 
\multicolumn{10}{l}{\small {Effelsberg, and Tidbinbilla.}} 
\endfoot
%
%
\end{longtable}

\begin{table}
\tbl{Column density of the known NLS1 masers}{%
\begin{tabular}{lrr} 
\hline
Source~~~~~~~~~~~&\nhh$^{\ast}$&   References~~~~~~ \\
&(10$^{22}$ cm$^{-2}$)&   \\
\hline
IRAS\,03450+0055 & --- $^{\dagger}$     &  \citet{rus96} \\
NGC\,4051  &   8 -- 37     &  \citet{mad06}    \\
Mrk\,766    &  20 -- 30 $^{\ddagger}$ $>$   &  \citet{ris11}   \\
NGC\,5506  &  3.40      &  \citet{mol13} \\
IGR\, J16385-2057  &  0.12      &         \citet{mol13}   \\
\hline
\end{tabular}}\label{tab2}
\begin{tabnote}
\footnotemark[$^{\ast}$] \nh are adopted from published X-ray data from IRAS 03450+0055 (ROSAT),
Mrk 766 (BeppoSAX), NGC 5506 and IGR J16385-2057 (INTEGRAL), and NGC 4051 (Chandra). 
\par\noindent
\footnotemark[$^{\dagger}$] Galactic \nh (soft X-ray) from the ROSAT All-Sky Survey is 5.89 $\times$ 10$^{20}$ cm$^{-2}$. 
\par\noindent
\footnotemark[$^{\ddagger}$] Lower limit value.
\end{tabnote}
\end{table}

\begin{center}
\begin{figure}[htbp]
\includegraphics[width=11cm]{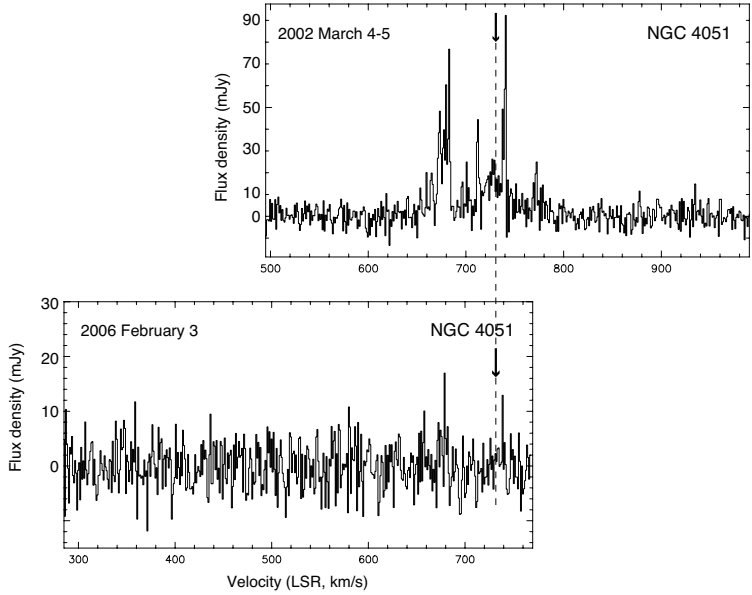}
\caption{Spectra of \ho maser ($\Delta\nu$ = 2.1 \kmss) between 275 \kms -- 775 \kms toward the centre of NGC\,4051, obtained with the Effelsberg 100 m in 2002 and 2006.  A vertical dotted line and arrows denote the systemic velocity of the galaxy (730 \kmss). The vertical axis denotes flux density scaled in Jansky, and the horizontal axis LSR velocity.}
\label{f1}
\end{figure}
%
\begin{figure}[htbp]
\includegraphics[width=10.5cm]{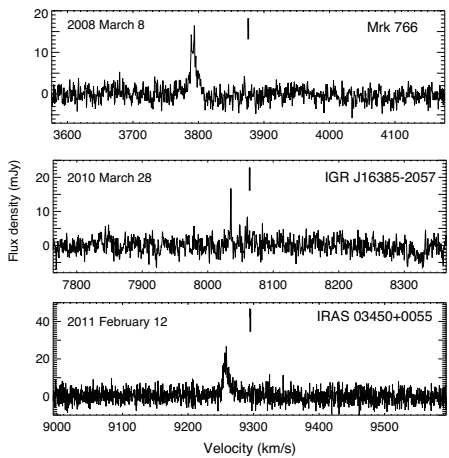}
\caption{Published GBT spectra of \ho masers in the 3 NLS1 galaxies, Mrk\,766, IGR\,J16385--2057, and IRAS\,03450+0055 \citep{tar11}. In these figures, heliocentric velocity definition 
is adopted and vertical bars denote systemic velocity of the galaxies.}
\label{f2}
\end{figure}
\end{center}
\clearpage


\begin{thebibliography}{}
%
\bibitem[Antonucci \&  Miller(1985)]{ant85}
 Antonucci, R. R. J.  \& Miller, J. S. \ 1985, \apj, 297, 621
\bibitem[Boller et al.(1996)]{bol96}  Boller, Th., Brandt, W. N., \& Fink, H. \ 1996, \aap, 305, 53
\bibitem[Boroson \& Green(1992)]{bor92}Boroson, T. A., Green, R. F.   \ 1992, \apjs, 80, 109
\bibitem[Braatz et al.(1996)]{bra96} Braatz, J. A., Wilson, A. S., \& Henkel, C. \ 1996, \apjs, 106, 51
\bibitem[Braatz et al.(1997)]{bra97} Braatz, J. A., Wilson, A. S., \& Henkel, C. \ 1997, \apjs, 110, 321
%
\bibitem[Braatz et al.(2004)]{bra04} Braatz, J. A., Henkel, C., Greenhill, L. J., Moran, J. M.,  \& Wilson, A. S. \ 2004, \apj, 617, L29
\bibitem[Braatz et al.(2008)]{bra08} Braatz, J. A., Gugliucci, N. E. \ 2008, \apj, 678, 96
\bibitem[Breen et al.(2013)]{bre13}Breen, S. L., Lovell, J. E. J., Ellingsen, S. P., Horiuchi, S.,  Beasley, A. J., Marvel, K.   \ 2013, \mnras, 432, 1382
\bibitem[Castangia et al.(2013)]{cas13} Castangia, P., Panessa, F., Henkel, C., Kadler, M., Tarchi, A . \ 2013, \aap, 308, 592
\bibitem[Claussen \&  Lo(1986)]{cla86}Claussen, M. J., \& Lo, K.-Y. \ 1986, \apj, 308, 592
\bibitem[Claussen et al.(1998)]{cla98} Claussen, M. J., Diamond, P. J.,  Braatz, J. A.,  Wilson, A. S.,  Henkel, C. \  1998, \apjl, 500, 129
\bibitem[Deo et al.(2006)]{deo06} Deo, R. P., Crenshaw, D. M.,  \& Kraemer, S. B. \ 2006, \aj, 132, 321
\bibitem[Doi et al.(2013)]{doi13} Doi, A., Asada, K., Fujisawa, K., Nagai, H., Hagiwara,  \& Y., Wajima, K.  \ 2013, \apj, 765, 69
\bibitem[Doi et al.(2015a)]{doi15a} Doi, A. \ 2015a, \pasj, 67, 157
\bibitem[Doi et al.(2015b)]{doi15b} Doi, A., Wajima, K., Hagiwara, Y.,  \& Inoue, M. \ 2015b, \apj, 798, L30
\bibitem[Gallimore et al.(2001)]{gal01} Gallimore J. F., Henkel C., Baum S. A., Glass I. S., Claussen M. J., Prieto M. A., Von Kap-herr A. 2001, \apj, 556, 694
\bibitem[Gao et al.(2017)]{gao17}Gao, F., Braatz, J. A., Reid, M. J., Condon, J. J., et al. 2017, \apj, 834, 52
\bibitem[Goodrich (1989)]{goo89} Goodrich, R. W. 1989, \apj, 342, 224
\bibitem[Greenhill et al.(1996)]{linc96} Greenhill, L. J., Gwinn, C. R., Antonucci, R., \& Barvainis, R. 1996, \apj, 472, L21
\bibitem[Greenhill et al.(2003a)]{linc03a} Greenhill, L. J.,  Kondratko, P. T., Lovell, J. E. J., Kuiper, T. B. H., Moran, J. M., Jauncey, D. L.,  Baines, G. P. \ 2003a, \apj, 582, L11
\bibitem[Greenhill et al.(2003b)]{linc03b} Greenhill, L. J., Booth, R. S.,  Ellingsen, S. P.,  Herrnstein, J. R., Jauncey, D. L., et al. \ 2003b, \apj, 590, 162
%
%
\bibitem[Greenhill et al.(2008)]{linc08} Greenhill, L. J., Tilak, A.,  \& Madejski, G. \ 2008, \apj, 686, L13
%
\bibitem[Greenhill et al.(2009)]{linc09} Greenhill, L. J., Kondratko, P. T., Moran, J. M.,  \& Tilak, A. \ 2009, \apj, 707, 787
%
\bibitem[Grupe \& Mathur (2004)]{gru04} Grupe \& Mathur \ 2004, \apjl, 606, 41
%
\bibitem[Hagiwara et al.(2001)]{hagi01} Hagiwara, Y., Diamond, P. J., Nakai, N.,  \& Kawabe, R. \ 2001, \apj, 560, 119
%
\bibitem[Hagiwara et al.(2002)]{hagi02} Hagiwara, Y., Diamond, P. J.,  \& Miyoshi, M. \ 2002, \aap, 383, 65
\bibitem[Hagiwara et al.(2003)]{hagi03} Hagiwara, Y., Diamond, P. J., Miyoshi, M., Rovilos, E., \& Baan, W. \ 2003, \mnras, 344, L53
%
\bibitem[Hagiwara(2007)]{hagi07} Hagiwara, Y. \ 2007, \aj, 133, 1176
%
\bibitem[Hagiwara \& Edwards(2015)]{hagi15} Hagiwara, Y., \& Edwards, P. G. \ 2015, \apj, 815, 124
%
\bibitem[Herrnstein et al.(1999)]{her99} Herrnstein, J. R., Moran, J. M., Greenhill, L. J., Diamond, P. J., et al., \ 1999, \nat, 400, 539
\bibitem[Ishihara et al.(2001)]{ish01} Ishihara, Y., Nakai, N., Iyomoto, N., Makishima, K., et al. 2001, \pasj, 53, 215
\bibitem[Komossa et al.(2006)]{kom06}  Komossa, S., Voges, W., Xu, D., Mathur, S., Adorf, H-.M., Lemson, G., Duschl, W. J., Grupe, D. \ 2006, \aj, 132, 531
\bibitem[Komossa et al.(2008)]{kom08}  Komossa, S., Xu, D., Zhou, H., Storchi-Bergmann, T., \& Binette, L. \ 2008, \apj, 680, 926
\bibitem[Kondratko et al.(2005)]{kon05}  Kondratko, P. T., Greenhill, L. J.,  \& Moran, J. M. \ 2005, \apj, 618, 618
\bibitem[Kondratko et al.(2006)]{kon06a}  Kondratko, P. T., Greenhill, L. J., Moran, J. M., Lovell, J. E. J., Kuiper, T. B. H.  et al., \ 2006, \apj, 638, 100
\bibitem[Kondratko et al.(2008)]{kon08}  Kondratko, P. T., Greenhill, L. J.,  \& Moran, J. M. \ 2008, \apj, 678, 87
\bibitem[Kuo et al.(2011)]{kuo11} Kuo, C. Y., Braatz, J. A., Condon, J. J.,  Impellizzeri,  C. M. V.,  Lo, K. Y.,  Zaw, I.,  Schenker, M.,  Henkel, C.,  Reid, M. J.,  Greene, J. E.  \ 2011, \apj, 727, 20
\bibitem[Madejski et al.(2006)]{mad06} Madejski, G., Done, C.,  \.{Z}ycki, P. T.,  \& Greenhill, L. \ 2006, \apj, 636, 75
\bibitem[Mamyoda et al.(2009)]{mam09}  Mamyoda, K.,  Nakai, N., Yamauchi, A., Diamond, P.,  et al.  2009, \pasj, 61, 1143
%
\bibitem[Masini et al.(2016)]{mas16} Masini, A., Comastri, A., Balokovi\'{c}, M., et al. \ 2016, \aap, 589, 59
%
\bibitem[Mathur et al.(2001)]{mat01} Mathur, S. Kuraszkiewicz, J. Czerny, B. \ 2001, New Astronomy, 6, 321
%
\bibitem[McHardy et al.(1995)]{mch95} McHardy,  I. M., Green, A. R., Done, C., et al. \ 1995, \mnras, 273, 549
\bibitem[Mineshige et al.(2000)]{mine00} Mineshige, S., Kawaguchi, T., Takeuchi, M., Hayashida, K. \ 2000, \pasj, 52, 499
%
\bibitem[Miyoshi et al.(1995)]{miyo95}
Miyoshi, M.  Moran, J., Herrnstein, J., Greenhill, L., Nakai, N., Diamond, P.,  \& Inoue, M. \ 1995, \nat, 373, 127
\bibitem[Molina et al.(2013)]{mol13} Molina, M., Bassani, L., Malizia, A., et al. \ 2013, \mnras, 433, 1687
%
\bibitem[Mullaney and Ward(2008)]{mul08}Mullaney, J. R. and Ward, M. J. \ 2008, \mnras, 385, 53
%
\bibitem[Mundell et al.(1995)]{mun95} Mundell, C. G., Pedlar,  A.,  Baum, S. A.,  O'Dea, C. P., Gallimore, J. F.,  \& Brinks, E.  \ 1995, \mnras, 272, 355
%
\bibitem[Nagar et al.(2002)]{nag02}  Nagar, N. M., Oliva, E., Marconi, A., Maiolino, R. \ 2002,  \aap, 391, L21
\bibitem[Osterbrock \& Pogge(1985)]{ost85} Osterbrock, D. E.,  \& Pogge, R. W.  \ 1985, \apj, 297, 166
%
\bibitem[Ott et al.(2013)]{ott13} Ott, J., Meier, D. S., McCoy, M., Peck, A.,  Impellizzeri, V., et al.  \ 2013, \apj, 771, L41
%
\bibitem[Panessa et al.(2011)]{pan11} Panessa, F., de Rosa, A., Bassani, L., Bazzano, A., Bird, A., Landi, R., Malizia, A., Miniutti, G., Molina, M., Ubertini, P. \ 2011, \mnras, 417, 2426
%
\bibitem[Peterson et al.(2000)]{pet00} Peterson, B. M., et al. 2000, ApJ, 542, 161
\bibitem[Reid et al.(2009)]{rei09}  Reid, M. J., Braatz, J. A., Condon, J.J.,  Greenhill, L. J., et al. \ 2009, \apj, 695, 287
\bibitem[Risaliti et al.(2011)]{ris11} Risaliti, G., Nardini, E., Salvati, M., et al.  \ 2011, \mnras, 410, 1027
\bibitem[Rush et al.(1996)]{rus96} Rush, B., Malkan, M.  A., Fink, H. H.,  \& Voges, W. \ 1996, \apj, 471, 190
%
\bibitem[Surcis et al.(2009)]{sur09} Surcis, G., Tarchi, A., Henkel, C., Ott, J., Lovell, J.,  \& Castangia, P.  \ 2009, \aap, 502, 529
%
\bibitem[Tarchi et al.(2011)]{tar11} Tarchi, A., Castangia, P., Columbano, A., Panessa, F.,  \& Braatz, J. A.  \ 2011, \aap, 532, 125
\bibitem[Turner et al.(2006)]{tur06} Turner, T. J.,  Miller, L.,  George, I. M.,  \& Reeves, J. N. \ 2006, \aap, 445, 59
%
\bibitem[Ulvestad et al.(1995)]{ulv95} 
Ulvestad, J. S., Antonucci, R. R. J., Goodrich, R. W. \ 1995, \apj, 109, 81
%
\bibitem[Urry  \&  Padovani(1995)]{urr95}
Urry, C. M.,  \& Padovani, P.  \ 1995, PASP, 107, 803
%
\bibitem[van den Bosch et al.(2016)]{bos16} van den Bosch, R. C. E., Greene, J. E., Braatz, J. A.,  Constantin, A.,  \& Kuo, C.-Yu. \ 2016, \apj, 819, 11
%
\bibitem[V\'{e}ron-Cetty et al.(2001)]{veron01} V\'{e}ron-Cetty, M.-P., V\'{e}ron,  P.,  \& Gon\c{c}alves, A. C.  \ 2001, \aap, 372, 730
%
\bibitem[Wagner(2013)]{wag13} Wagner, A.  \ 2013, \aap, 560, 12
%
\bibitem[Wang \& Lu (2001)]{wan01} Wang, T.,  \& Lu,  Y.  \ 2001, \aap, 377, 52
%
\bibitem[Whalen et al.(2006)]{wha06}Whalen, D. J., Laurent-Muehleisen, S. A., Moran, E. C.,  \& Becker, R. H.  \ 2006, \aj, 131, 1948
%
\bibitem[Woo et al.(2015)]{woo15}Woo, J.-H., Yoon, Y., Park, S., Park, D., Kim, S. C. \ 2015, \apj, 801, 38
%
\bibitem[Zhang et al.(2006)]{zhan06} Zhang, J.~S., Henkel, C., Kadler, M., Greenhill, L. J.,  Nagar, N.,  Wilson, A.~S., \& Braatz, J. A. \ 2006, \aap, 450, 933
\bibitem[Zhu et al.(2011)]{zhu11}  Zhu, G., Zaw, I., Blanton, M. R., Greenhill, L. J. \ 2011, \apj, 742, 73
\end{thebibliography}
\end{document}